\documentclass[11pt,a4paper,twocolumn]{article}

\usepackage[utf8]{inputenc}
\usepackage[T1]{fontenc}
\usepackage{mathptmx}           
\usepackage[margin=2cm]{geometry}
\usepackage{graphicx}
\usepackage{booktabs}
\usepackage{tabularx}
\usepackage{xcolor}
\usepackage{hyperref}
\usepackage{url}
\usepackage{cite}

\usepackage{enumitem}
\usepackage{tikz}
\usepackage{float}
\usepackage{multirow}
\usepackage{array}
\usepackage{caption}
\usepackage{balance}

\usetikzlibrary{shapes.geometric, arrows.meta, positioning, fit, backgrounds, calc}

\hypersetup{
  colorlinks=true,
  linkcolor=blue!60!black,
  citecolor=blue!60!black,
  urlcolor=blue!60!black,
  pdfauthor={Jonathan Shelby},
  pdftitle={Architectural Implications of the UK Cyber Security and Resilience Bill},
  pdfsubject={Security Architecture, Zero Trust, CS\&R Bill}
}

\setlist{nosep,leftmargin=*}

\title{\Large\bfseries Architectural Implications of the UK Cyber Security\\and Resilience Bill: A Practitioner's Guide to\\Security Architecture Transformation}

\author{%
  \normalsize Jonathan Shelby\\
  \small Department of Computer Science, University of Oxford\\
  \small Hertford College, Oxford, UK\\
  \small\texttt{jonathan.shelby@cs.ox.ac.uk}
}

\date{\small April 2026}

\begin{document}

\twocolumn[
  \maketitle
  \begin{@twocolumnfalse}
  \begin{abstract}
  \noindent
  The UK Cyber Security and Resilience (CS\&R) Bill represents the most significant reform of UK cyber legislation since the Network and Information Systems (NIS) Regulations 2018. While existing analysis has addressed the Bill's regulatory requirements, there is a critical gap in guidance on the \emph{architectural} implications for organisations that must achieve and demonstrate compliance. This paper argues that the CS\&R Bill's provisions---expanded scope to managed service providers (MSPs), data centres, and critical suppliers; mandatory 24/72-hour dual incident reporting; supply chain security duties; and Secretary of State powers of direction---collectively constitute an architectural forcing function that renders perimeter-centric and point-solution security postures structurally non-compliant. We present a systematic mapping of the Bill's key provisions to specific architectural requirements, demonstrate that Zero Trust Architecture (ZTA) provides the most coherent technical foundation for meeting these obligations, and propose a reference architecture and maturity-based adoption pathway for CISOs and security architects. The paper further addresses the cross-regulatory challenge facing UK financial services firms operating under simultaneous CS\&R, DORA, and NIS2 obligations, and maps the architectural framework against the NCSC Cyber Assessment Framework v4.0. This work extends a companion practitioner guide to the Bill~\cite{shelby2025csr} by translating regulatory analysis into actionable architectural strategy.

  \end{abstract}
  \vspace{1em}
  \end{@twocolumnfalse}
]

\section{Introduction}
\label{sec:introduction}

The UK Cyber Security and Resilience (Network and Information Systems) Bill~\cite{ukgov2025csrbill}, introduced to Parliament on 12 November 2025, represents the most significant overhaul of UK cross-sector cyber legislation since the original NIS Regulations came into force in 2018~\cite{ukgov2018nis}. Having completed its second reading in January 2026 and progressed through committee stage in February, the Bill is expected to receive Royal Assent later in 2026, with phased implementation extending through to 2028~\cite{ukgov2026factsheets}.

The scale of the threat driving this reform is substantial. In the year preceding September 2025, the National Cyber Security Centre (NCSC) managed 429 cyber incidents, of which 204 were nationally significant---more than double the previous year~\cite{ncsc2025annualreview}. Over 40\% of UK businesses experienced cyber attacks in 2024, with estimated annual costs of \pounds14.7 billion, equivalent to 0.5\% of GDP~\cite{ukgov2025cyberbreaches, ukgov2025cyberimpact}. An independent assessment found that 95\% of UK critical national infrastructure organisations experienced a data breach in 2024~\cite{bridewell2025cni}.

A companion paper by the present author~\cite{shelby2025csr} provides a comprehensive practitioner's guide to the Bill's regulatory requirements, including sector-by-sector compliance roadmaps, a dual-compliance framework for organisations subject to both CS\&R and the EU Digital Operational Resilience Act (DORA)~\cite{eu2022dora}, and detailed gap analysis tools. That work addresses the question: \emph{what does the Bill require?} The present paper addresses the question that naturally follows: \emph{what does compliant security architecture actually look like?}

This distinction matters because the CS\&R Bill is not merely a compliance exercise that can be satisfied through policy documentation and annual audits. Its provisions---particularly the expanded scope to managed service providers and critical suppliers, the 24/72-hour incident reporting regime, and the Secretary of State's powers of direction---impose architectural requirements that cannot be met without fundamental changes to how organisations design, deploy, and govern their security infrastructure.

The central argument of this paper is that the CS\&R Bill functions as an \emph{architectural forcing function}: its cumulative requirements make perimeter-centric and point-solution security architectures structurally non-compliant, and drive organisations toward Zero Trust Architecture (ZTA)~\cite{nist800207} as the coherent technical response. This is not because ZTA is fashionable, but because the Bill's obligations---continuous assurance, supply chain visibility, rapid detection and reporting, cross-boundary trust management---are functionally ZTA requirements expressed in regulatory language.

The paper is structured as follows. Section~\ref{sec:background} provides essential context on the Bill and the current state of UK enterprise security architecture. Section~\ref{sec:gap} identifies the specific architectural gap that the Bill exposes. Section~\ref{sec:mapping} presents a systematic mapping of Bill provisions to architectural requirements. Section~\ref{sec:zta} makes the case for ZTA as the coherent architectural response. Section~\ref{sec:reference} proposes a reference architecture. Section~\ref{sec:supply} addresses supply chain architecture. Section~\ref{sec:incident} covers incident detection and reporting architecture. Section~\ref{sec:crossreg} examines cross-regulatory alignment. Section~\ref{sec:caf} maps the framework against NCSC CAF v4.0. Section~\ref{sec:roadmap} presents a maturity-based adoption roadmap. Section~\ref{sec:conclusion} concludes with recommendations for CISOs and security architects.

\section{Background}
\label{sec:background}

\subsection{The CS\&R Bill: Key Provisions}

The Bill operates through three pillars of reform: expanded scope, effective regulators, and enabling resilience~\cite{ukgov2026factsheets}. For the purposes of architectural analysis, the following provisions are most consequential.

\textbf{Expanded scope.} The Bill brings into the NIS regulatory perimeter three new categories of entity: data centres (regulated by Ofcom, threshold of medium/large or enterprise data centres meeting defined criteria); managed service providers (MSPs, regulated by the ICO, covering organisations providing third-party IT services); and large load controllers managing electrical load for smart appliances~\cite{ukgov2026factsheets}. Additionally, regulators gain the power to designate \emph{critical suppliers}---organisations whose cyber vulnerability could severely affect the essential services they supply---imposing direct regulatory obligations on them for the first time.

\textbf{Incident reporting.} The Bill introduces a mandatory dual-notification regime: initial notification within 24 hours and a fuller report within 72 hours of becoming aware of a qualifying incident. The scope of reportable incidents is expanded beyond significant service disruption to include incidents with the \emph{potential} to cause significant impacts---a critical distinction that requires detection capabilities beyond those needed for post-disruption reporting. Data centres, digital service providers, and MSPs must additionally inform affected customers~\cite{ukgov2026factsheets}.

\textbf{Supply chain duties.} Secondary legislation is expected to impose duties on operators of essential services (OESs) and relevant digital service providers (RDSPs) to manage cyber risk across their supply chains through ``appropriate and proportionate measures,'' potentially including contractual requirements, security checks, and business continuity planning.

\textbf{Powers of direction.} The Secretary of State gains powers to direct both regulators and regulated entities to take targeted, proportionate action in response to threats that risk UK national security. This power creates an architectural requirement for organisations to be able to implement security changes rapidly in response to government direction.

\textbf{Cost recovery and enforcement.} Regulators gain full cost recovery powers and enhanced enforcement capabilities, with maximum financial penalties aligned to comparable regimes such as GDPR (up to \pounds17.5 million or 4\% of global turnover).

\subsection{Current State of UK Enterprise Security Architecture}

The majority of UK enterprises, particularly those in sectors already regulated under the NIS Regulations 2018, operate security architectures that are fundamentally perimeter-centric in design~\cite{kindervag2010zt}. While many have adopted individual Zero Trust technologies---multi-factor authentication, identity-aware proxies, endpoint detection and response---these are typically deployed as point solutions overlaid on an implicit-trust network model rather than as components of a coherent architectural paradigm.

This creates several structural weaknesses relevant to CS\&R compliance:

\emph{Boundary-dependent trust models.} Organisations implicitly trust traffic and users within the corporate network perimeter and concentrate security controls at ingress/egress points. This model fails when MSPs, cloud services, and supply chain partners require authenticated access deep into organisational systems.

\emph{Detection latency.} Perimeter-centric architectures typically lack the continuous monitoring and per-transaction verification capabilities needed to detect incidents within the timeframes demanded by 24-hour initial notification. Mean time to detect remains measured in days or weeks across many UK sectors.

\emph{Siloed security functions.} Network security, identity and access management, endpoint protection, and application security are frequently operated as independent capabilities with limited integration, producing fragmented visibility that undermines both detection and reporting.

\emph{Supply chain opacity.} Existing architectures rarely provide granular visibility into third-party access patterns, data flows, and risk posture---a fundamental gap given the Bill's supply chain provisions.

\section{The Architectural Gap}
\label{sec:gap}

The CS\&R Bill exposes a structural mismatch between what the legislation demands and what prevailing security architectures can deliver. This gap is not merely a matter of deploying additional controls; it is an architectural problem that requires rethinking the foundations upon which security is built.

\subsection{From Compliance Checkboxes to Continuous Assurance}

Under the existing NIS Regulations, compliance has often been demonstrated through periodic assessment: annual audits, penetration tests, and self-assessment questionnaires. The CS\&R Bill's provisions, particularly the expanded incident reporting requirements and the Secretary of State's powers of direction, shift the expectation from periodic compliance to continuous assurance. Organisations must be able to demonstrate---at any point in time, and potentially at short notice---that their security posture meets the required standard and that they can detect and report incidents within mandated timeframes.

This shift has profound architectural implications. A security architecture designed around annual compliance cycles optimises for documentation and point-in-time evidence. A security architecture designed for continuous assurance optimises for real-time visibility, automated policy enforcement, and dynamic response---fundamentally different design objectives that produce fundamentally different architectural decisions.

\subsection{The MSP and Supply Chain Challenge}

The inclusion of MSPs and the critical supplier designation power create an architectural challenge that perimeter-centric models cannot address. When an organisation's essential service depends on an MSP that has deep access to its network and information systems, the security of that service is a function of the combined security posture of both entities. The traditional approach---a firewall between the organisation and its MSP, supplemented by contractual security requirements---is insufficient when the MSP has legitimate administrative access to systems that, if compromised, would directly affect service delivery.

The Capita incident of March 2023~\cite{capita2023}, the Advanced/NHS incident of August 2022~\cite{advanced2022nhs}, and the Synnovis/NHS pathology attack of June 2024~\cite{synnovis2024} all illustrate this failure mode: a supplier's compromise propagated directly to the essential services it supported, precisely because the architectural boundary between supplier and consumer was either absent or insufficiently granular.

\subsection{Architectural Lessons from UK Incidents}

The case for architectural transformation is not theoretical. Three recent UK incidents illustrate the failure modes that the CS\&R Bill is designed to address, and each points to specific architectural deficiencies.

The \textbf{Capita incident} (March 2023)~\cite{capita2023} compromised a managed service provider with contracts across UK government, local authorities, and major pension funds. The attack propagated through Capita's internal systems to client-facing services, disrupting multiple public services simultaneously. The architectural lesson is that Capita's clients had no meaningful visibility into, or architectural isolation from, their service provider's internal security posture. A Zero Trust supply chain architecture---with brokered access, session-level authorisation, and continuous monitoring of MSP activity---would have both limited the blast radius and provided earlier detection.

The \textbf{Advanced/NHS incident} (August 2022)~\cite{advanced2022nhs} saw a ransomware attack on a managed software provider disrupt NHS 111 services, mental health trusts, and other healthcare organisations across England. The attacker gained initial access through a legitimate account that lacked multi-factor authentication. The incident exposed two architectural failures: the absence of continuous identity verification (a fundamental ZTA requirement) and the lack of segmentation between Advanced's internal systems and the NHS-facing services they supported.

The \textbf{Synnovis/NHS pathology incident} (June 2024)~\cite{synnovis2024}---cited in the Bill's own supporting documentation---resulted in over 11,000 postponed appointments and procedures and contributed to the death of a patient. A single supplier's vulnerability directly affected patient safety in multiple hospitals. This incident was the catalytic example for the Bill's critical supplier designation power and underscores the need for architectural controls that treat supplier access as a continuously verified, granularly controlled, and comprehensively monitored trust relationship.

In each case, the prevailing architecture treated the supplier relationship as a contractual and perimeter problem rather than an architectural one. The CS\&R Bill's provisions are a direct response to this pattern of failure.

\subsection{The 24-Hour Detection Problem}

The Bill's requirement for initial notification within 24 hours of becoming aware of a qualifying incident---including incidents with the \emph{potential} to cause significant impact---creates a detection latency constraint that many current architectures cannot satisfy. The Cyber Security Breaches Survey 2025~\cite{ukgov2025cyberbreaches} indicates that a significant proportion of organisations lack the continuous monitoring capabilities, log aggregation infrastructure, and automated correlation needed to identify qualifying incidents within this timeframe.

This is not a tooling problem but an architectural one. Effective sub-24-hour detection requires: comprehensive telemetry across all components of the network and information systems relied upon for the essential service; centralised log aggregation with sufficient retention and query performance; automated correlation and alerting calibrated to the Bill's significance thresholds; and integration between detection systems and the reporting workflow. These capabilities must be designed into the architecture, not bolted on after the fact.

\section{Mapping Provisions to Architecture}
\label{sec:mapping}

Table~\ref{tab:mapping} presents a systematic mapping of the CS\&R Bill's key provisions to specific architectural requirements. This mapping forms the foundation for the reference architecture proposed in Section~\ref{sec:reference}.

\begin{table*}[t]
\centering
\caption{Mapping of CS\&R Bill Provisions to Architectural Requirements}
\label{tab:mapping}
\small
\begin{tabularx}{\textwidth}{>{\raggedright\arraybackslash}p{3.2cm}>{\raggedright\arraybackslash}p{4.5cm}>{\raggedright\arraybackslash}X}
\toprule
\textbf{Bill Provision} & \textbf{Obligation} & \textbf{Architectural Requirement} \\
\midrule
Expanded scope: MSPs & MSPs must meet security duties; customers remain accountable for service delivery & Identity federation with MSPs; granular access control for third-party administrative access; continuous monitoring of MSP activity; mutual attestation \\
\addlinespace
Expanded scope: Data centres & Data centres as essential services; Ofcom as regulator & Infrastructure-layer security controls; physical-logical convergence; tenant isolation architecture; environmental monitoring integration \\
\addlinespace
Critical supplier designation & Regulators may designate critical suppliers subject to NIS duties & Supply chain risk management architecture; real-time supplier risk scoring; automated assurance data collection; contractual security architecture requirements \\
\addlinespace
24-hour initial notification & Report qualifying incidents within 24 hours of awareness & Comprehensive telemetry; centralised SIEM/SOAR; automated significance classification; integrated reporting workflows; sub-24-hour mean-time-to-detect target \\
\addlinespace
72-hour full report & Detailed incident report within 72 hours & Forensic readiness architecture; immutable logging; evidence preservation; structured incident data model \\
\addlinespace
Customer notification (MSPs/DCs) & Inform affected customers of incidents & Cross-boundary incident communication; customer impact assessment capability; automated notification infrastructure \\
\addlinespace
Powers of direction & Secretary of State may direct specific security actions & Architecturally agile security controls; API-driven policy enforcement; rapid configuration change capability; pre-validated response playbooks \\
\addlinespace
Supply chain duties & OESs/RDSPs must manage supplier cyber risk & Zero Trust supply chain integration; supplier access segmentation; continuous third-party assurance; data flow mapping across supply chain \\
\addlinespace
Cost recovery & Regulators may recover costs from regulated entities & Audit-ready architecture; automated compliance evidence generation; structured reporting interfaces to regulators \\
\addlinespace
CAF alignment & NCSC CAF v4.0 as assurance framework & Architecture mapped to CAF objectives (A--D); control traceability; continuous assessment capability \\
\bottomrule
\end{tabularx}
\end{table*}

Several patterns emerge from this mapping.

First, virtually every provision generates a requirement for continuous, rather than periodic, security capability. Second, the cross-boundary provisions (MSPs, critical suppliers, customer notification) demand security architectures that extend trust and verification beyond the organisational perimeter. Third, the detection and reporting timeline constraints require architectural integration between telemetry, analytics, and workflow systems that is inconsistent with siloed security functions. Fourth, the powers of direction create a unique requirement for architectural agility---the ability to implement directed security changes rapidly without destabilising the operational environment.

\begin{figure*}[t]
\centering
\begin{tikzpicture}[
  prov/.style={draw=red!70!black, fill=red!20, rounded corners=4pt, minimum width=2.8cm, minimum height=0.9cm, font=\small\bfseries, align=center, text=red!80!black, line width=0.6pt},
  breaks/.style={draw=gray!60, fill=gray!15, rounded corners=4pt, minimum width=3.2cm, minimum height=0.9cm, font=\small, align=center, text=gray!70!black, line width=0.6pt},
  resp/.style={draw=teal!70!black, fill=teal!20, rounded corners=4pt, minimum width=3.2cm, minimum height=0.9cm, font=\small\bfseries, align=center, text=teal!80!black, line width=0.6pt},
  arr/.style={-{Stealth[length=5pt, width=4pt]}, line width=0.8pt, gray!70},
  concl/.style={draw=blue!60!purple, fill=blue!12!purple!12, rounded corners=6pt, minimum width=9cm, minimum height=1cm, font=\small\bfseries, align=center, text=blue!60!purple, line width=0.8pt}
]
\node[font=\footnotesize\itshape, text=gray!60] at (0, 0.6) {Bill provision};
\node[font=\footnotesize\itshape, text=gray!60] at (4.5, 0.6) {What it breaks};
\node[font=\footnotesize\itshape, text=gray!60] at (9.5, 0.6) {Architectural response};

\node[prov] (p1) at (0, 0) {MSP regulation};
\node[prov] (p2) at (0, -1.2) {24hr reporting};
\node[prov] (p3) at (0, -2.4) {Supply chain duty};
\node[prov] (p4) at (0, -3.6) {Critical suppliers};
\node[prov] (p5) at (0, -4.8) {Powers of direction};
\node[prov] (p6) at (0, -6.0) {Data centre scope};

\node[breaks] (b1) at (4.5, 0) {VPN full-network trust};
\node[breaks] (b2) at (4.5, -1.2) {Days-to-detect MTTD};
\node[breaks] (b3) at (4.5, -2.4) {Annual questionnaires};
\node[breaks] (b4) at (4.5, -3.6) {Contractual-only controls};
\node[breaks] (b5) at (4.5, -4.8) {Manual change processes};
\node[breaks] (b6) at (4.5, -6.0) {Shared-nothing tenancy};

\node[resp] (r1) at (9.5, 0) {ZTNA + PAM + MFA};
\node[resp] (r2) at (9.5, -1.2) {SIEM + SOAR pipeline};
\node[resp] (r3) at (9.5, -2.4) {Mutual attestation};
\node[resp] (r4) at (9.5, -3.6) {Supply chain ZTA};
\node[resp] (r5) at (9.5, -4.8) {API-driven policy};
\node[resp] (r6) at (9.5, -6.0) {Tenant isolation arch.};

\foreach \i in {1,...,6} {
  \draw[arr] (p\i) -- (b\i);
  \draw[arr] (b\i) -- (r\i);
}

\node[concl] (zta) at (4.75, -7.6) {Zero Trust Architecture is the coherent response};
\draw[arr, line width=1pt] (p6.south) |- ([yshift=-4pt]zta.west);
\draw[arr, line width=1pt] (r6.south) |- ([yshift=-4pt]zta.east);
\end{tikzpicture}
\caption{The CS\&R Bill as an architectural forcing function: each provision breaks a specific legacy assumption and demands an architectural response that converges on Zero Trust.}
\label{fig:forcing}
\end{figure*}
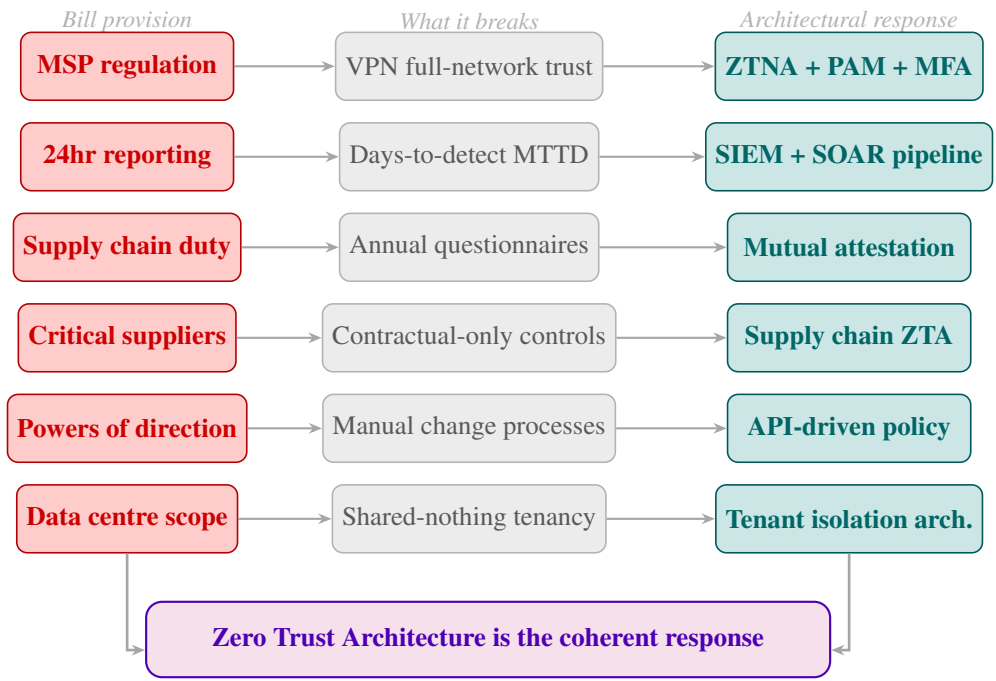

Figure~\ref{fig:forcing} illustrates this convergence. Each provision breaks a specific assumption embedded in prevailing security architectures, and the architectural responses consistently converge on Zero Trust principles.

\section{Zero Trust Architecture as the Coherent Response}
\label{sec:zta}

\subsection{Why Zero Trust Maps to CS\&R}

Zero Trust Architecture, as defined by NIST SP 800-207~\cite{nist800207}, is founded on the principle that trust is never granted implicitly based on network location, asset ownership, or physical/logical position. Instead, every access request is evaluated dynamically based on all available data points---identity, device health, behaviour, context, and risk---before granting the minimum necessary access for the minimum necessary duration.

The alignment between ZTA's foundational tenets and the CS\&R Bill's architectural requirements is not coincidental; both emerge from the same recognition that traditional perimeter-based security models fail in interconnected, supply-chain-dependent operational environments. Table~\ref{tab:zta_alignment} maps the core ZTA tenets to CS\&R provisions.

\begin{table}[h]
\centering
\caption{ZTA Tenets Mapped to CS\&R Provisions}
\label{tab:zta_alignment}
\small
\begin{tabularx}{\columnwidth}{>{\raggedright\arraybackslash}p{2.2cm}>{\raggedright\arraybackslash}X}
\toprule
\textbf{ZTA Tenet} & \textbf{CS\&R Alignment} \\
\midrule
Never trust, always verify & MSP/supplier access must be continuously authenticated and authorised \\
\addlinespace
Least privilege access & Critical supplier provisions require granular, role-based access control \\
\addlinespace
Assume breach & 24-hour detection requirement implies continuous monitoring and anomaly detection \\
\addlinespace
Continuous verification & Continuous assurance model replacing periodic compliance \\
\addlinespace
Micro-segmentation & Supply chain isolation and tenant separation in data centres \\
\addlinespace
Device and identity awareness & Powers of direction require rapid, granular enforcement capability \\
\bottomrule
\end{tabularx}
\end{table}

\subsection{ZTA as Architectural Strategy, Not Product}

A critical distinction for CISOs and security architects is that ZTA is an architectural strategy, not a product category. The market is saturated with vendors claiming ``Zero Trust'' capabilities, but deploying a Zero Trust Network Access (ZTNA) product does not constitute a Zero Trust Architecture any more than deploying a firewall constitutes a defence-in-depth strategy~\cite{gartner2024zt}.

For CS\&R compliance purposes, ZTA must be understood as a \emph{design philosophy} that governs decisions across all security domains: identity and access management, network architecture, endpoint security, application security, data protection, and security operations. The architectural value of ZTA in the CS\&R context lies in the \emph{coherence} it provides---a single set of design principles that simultaneously addresses MSP access control, supply chain visibility, incident detection, and regulatory reporting.

\subsection{Challenges and Limitations}

Adopting ZTA is not without challenges. Legacy operational technology (OT) environments---prevalent in energy, transport, and water sectors already under NIS regulation---often cannot support the continuous authentication and encryption that ZTA demands. Organisations must adopt a pragmatic, risk-based approach that applies ZTA principles where technically feasible while implementing compensating controls for legacy systems. The CISA Zero Trust Maturity Model~\cite{cisa2023ztmm} provides a useful framework for managing this transition, defining five pillars (Identity, Devices, Networks, Applications \& Workloads, and Data) across progressive maturity levels.

\subsection{The OT/IT Convergence Challenge}

The CS\&R Bill's expanded scope and enhanced security duties have particular implications for sectors that rely heavily on operational technology: energy, transport, drinking water, and increasingly, data centre infrastructure. These environments present unique architectural challenges for ZTA adoption that merit dedicated consideration.

\textbf{Protocol and legacy constraints.} Many OT systems operate on industrial protocols (Modbus, DNP3, IEC 61850, BACnet) that lack native support for encryption, authentication, or session-level authorisation. Implementing ZTA principles in these environments requires an overlay approach: deploying policy enforcement points at the boundary between IT and OT segments, implementing unidirectional security gateways where data must flow from OT to IT (e.g., for SIEM telemetry) without exposing OT systems to IT-originated traffic, and using protocol-aware inspection at segmentation boundaries.

\textbf{Availability primacy.} OT environments typically prioritise availability over confidentiality and integrity---the inverse of most IT security architectures. The CS\&R Bill's ``all-hazards'' approach recognises this, requiring risk management of both cyber and physical threats to network and information systems. The architectural response is not to impose IT-style ZTA on OT systems wholesale, but to apply ZTA principles to the \emph{trust boundaries} between OT and IT, ensuring that compromise of the IT environment cannot propagate to operational systems and vice versa. This boundary-focused approach satisfies the Bill's requirements while respecting OT operational constraints.

\textbf{Telemetry challenges.} Many OT systems generate limited security-relevant telemetry, creating potential gaps in the detection coverage needed for 24-hour incident notification. Compensating architectural patterns include passive network monitoring (protocol analysis without active probing), security-instrumented safety systems that can detect anomalous process behaviour, and data diodes that export OT telemetry to IT-hosted SIEM platforms without creating a return path.

Organisations in affected sectors should design their CS\&R compliance architecture with explicit OT/IT boundary treatment, applying full ZTA principles on the IT side while implementing proportionate, availability-preserving controls on the OT side with rigorous segmentation between the two domains.

\section{A Reference Architecture for CS\&R Compliance}
\label{sec:reference}

\begin{figure*}[t]
\centering
\begin{tikzpicture}[
  dbox/.style={rounded corners=6pt, minimum height=1.5cm, text width=4.2cm, align=center, font=\small, line width=0.6pt},
  bnd/.style={rounded corners=5pt, minimum height=1.2cm, text width=2.8cm, align=center, font=\small, line width=0.6pt},
  wide/.style={rounded corners=6pt, minimum height=1.2cm, text width=11.5cm, align=center, font=\small, line width=0.6pt},
  arr/.style={-{Stealth[length=4pt, width=3.5pt]}, line width=0.7pt, gray!65}
]
\draw[rounded corners=12pt, draw=gray!50, fill=gray!6, line width=0.8pt] (-0.5, 1.2) rectangle (12.0, -9.0);
\node[font=\small\bfseries, text=gray!80!black] at (5.75, 0.8) {CS\&R-Compliant Zero Trust Reference Architecture};

\node[dbox, draw=violet!70, fill=violet!18, text=violet!85!black] (d1) at (2.5, -0.2) {\textbf{Domain 1: Identity}\\{\footnotesize ABAC $\cdot$ PAM $\cdot$ Continuous auth}};
\node[dbox, draw=teal!70, fill=teal!18, text=teal!85!black] (d2) at (9.0, -0.2) {\textbf{Domain 2: Network}\\{\footnotesize Micro-seg $\cdot$ ZTNA $\cdot$ NDR}};
\node[dbox, draw=blue!60, fill=blue!15, text=blue!80!black] (d3) at (2.5, -2.3) {\textbf{Domain 3: Data}\\{\footnotesize Classification $\cdot$ DLP $\cdot$ Flow maps}};
\node[dbox, draw=orange!70!red, fill=orange!18!red!8, text=orange!80!red!80!black] (d4) at (9.0, -2.3) {\textbf{Domain 4: SecOps}\\{\footnotesize SIEM + SOAR $\cdot$ 24/72hr reporting}};
\node[wide, draw=green!60!black, fill=green!15, text=green!70!black] (d5) at (5.75, -4.2) {\textbf{Domain 5: Governance, Risk \& Compliance}\\{\footnotesize Continuous control monitoring $\cdot$ CAF v4.0 $\cdot$ Board reporting $\cdot$ CS\&R / DORA / NIS2}};

\draw[arr] (d1.east) -- (d2.west);
\draw[arr] (d1.south) -- (d3.north);
\draw[arr] (d2.south) -- (d4.north);
\draw[arr] (d3.south) -- ([xshift=-3.2cm]d5.north);
\draw[arr] (d4.south) -- ([xshift=3.2cm]d5.north);

\node[bnd, draw=orange!70, fill=orange!18, text=orange!80!black] (tb1) at (1.5, -5.9) {\textbf{MSP/Supplier}\\{\footnotesize Brokered access}};
\node[bnd, draw=orange!70, fill=orange!18, text=orange!80!black] (tb2) at (5.75, -5.9) {\textbf{OT/IT Boundary}\\{\footnotesize Unidirectional gw}};
\node[bnd, draw=orange!70, fill=orange!18, text=orange!80!black] (tb3) at (10.0, -5.9) {\textbf{Data Centre}\\{\footnotesize Tenant isolation}};

\draw[arr] ([xshift=-3.8cm]d5.south) -- (tb1.north);
\draw[arr] (d5.south) -- (tb2.north);
\draw[arr] ([xshift=3.8cm]d5.south) -- (tb3.north);

\node[wide, draw=purple!50!pink, fill=purple!10!pink!10, text=purple!70!pink!70!black] (caf) at (5.75, -7.7) {\textbf{NCSC Cyber Assessment Framework v4.0}\\{\footnotesize Objective A: Risk $\cdot$ B: Protect $\cdot$ C: Detect $\cdot$ D: Respond}};

\draw[arr] (tb1.south) -- ([xshift=-3.8cm]caf.north);
\draw[arr] (tb2.south) -- (caf.north);
\draw[arr] (tb3.south) -- ([xshift=3.8cm]caf.north);
\end{tikzpicture}
\caption{Five-domain Zero Trust reference architecture for CS\&R compliance, with cross-cutting trust boundaries and NCSC CAF v4.0 assurance mapping.}
\label{fig:refarch}
\end{figure*}
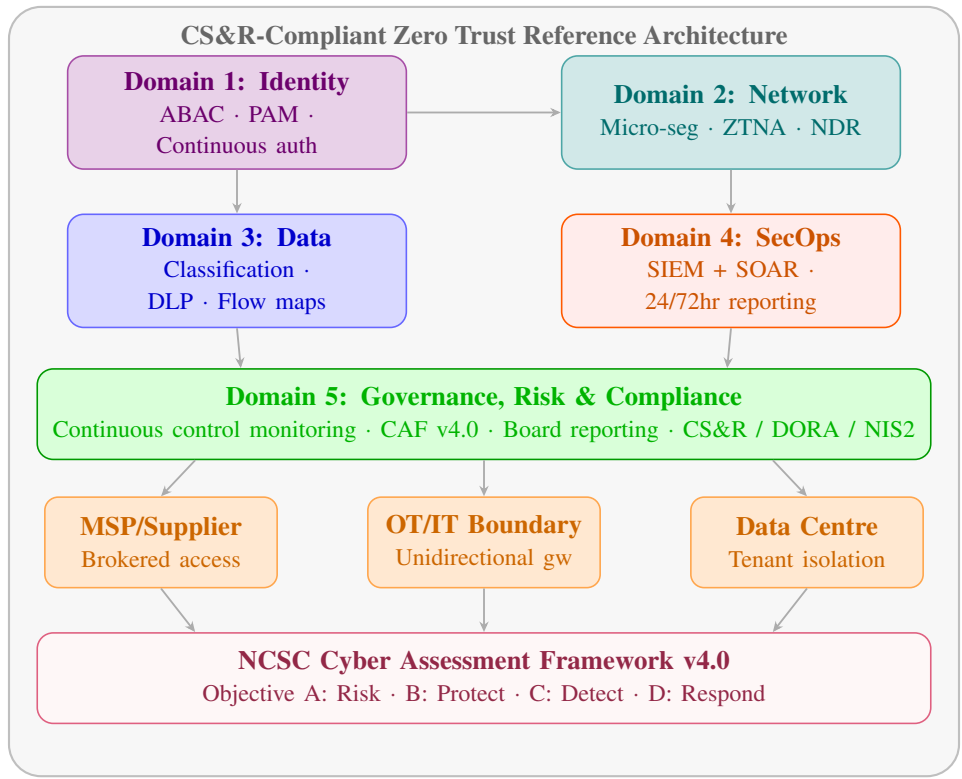

This section proposes a reference architecture structured around five architectural domains, each addressing specific CS\&R obligations while adhering to ZTA design principles. The architecture is intentionally technology-agnostic: it specifies capabilities rather than products, enabling organisations to implement using their existing or preferred technology stack. Figure~\ref{fig:refarch} provides a high-level view of the five domains, their relationships, and the cross-cutting trust boundaries they address.

\subsection{Domain 1: Identity and Access Governance}

Identity is the primary control plane in a CS\&R-compliant architecture. Every access decision---whether by internal users, MSP administrators, automated processes, or supply chain systems---must be mediated by an identity-aware policy engine that evaluates trust continuously.

The significance of this domain for CS\&R compliance cannot be overstated. The Bill's expanded scope brings MSPs under direct regulation, but it does not remove the responsibility of their clients for the security of their essential services. This creates a shared accountability model in which identity governance is the primary mechanism for managing risk across organisational boundaries. An MSP administrator with domain-level access to a client's Active Directory has, from a technical perspective, the same access as a malicious insider---and the architecture must treat them accordingly.

\textbf{Architectural requirements:}
\begin{itemize}
\item Unified identity store spanning internal users, MSP personnel, service accounts, and supply chain integrations, with federated identity where direct provisioning is impractical. The identity store must be authoritative: shadow IT accounts and local credentials outside the identity governance perimeter represent compliance gaps.
\item Policy engine implementing attribute-based access control (ABAC) with contextual risk signals: user identity, device posture, location, time, behaviour baseline, and resource sensitivity. The policy engine should be capable of expressing complex conditions---for example, granting an MSP engineer access to a specific database server only from a managed device, during UK business hours, with an active MFA session, and only for the duration of an approved change request.
\item Privileged access management (PAM) with session recording, just-in-time elevation, and time-bounded access grants for all administrative actions by MSPs and suppliers. Session recordings must be stored in an immutable store outside the MSP's administrative control---a critical forensic readiness requirement under the 72-hour reporting obligation.
\item Continuous authentication: session-level re-evaluation triggered by context changes (e.g., network transition, anomalous behaviour, device posture change). This goes beyond initial login MFA to continuous, risk-adaptive authentication throughout the session lifecycle.
\item Service account governance: automated discovery, rotation, and monitoring of non-human identities (service accounts, API keys, machine certificates) that provide system-to-system access. These accounts are frequently overlooked in identity programmes yet represent a significant proportion of the attack surface exploited in supply chain compromises.
\end{itemize}

\subsection{Domain 2: Network Architecture and Segmentation}

The network must evolve from a flat, implicitly trusted topology to a segmented, policy-enforced architecture in which access is granted per-session based on identity and context, not network location.

\textbf{Architectural requirements:}
\begin{itemize}
\item Micro-segmentation~\cite{shore2023microseg} isolating critical systems supporting essential services from general corporate infrastructure, with segment-to-segment traffic subject to policy enforcement.
\item Software-defined perimeter (SDP) or ZTNA for all remote and third-party access, eliminating VPN-based full-network access patterns.
\item East-west traffic inspection between segments, particularly for MSP-accessible segments, using network detection and response (NDR) capabilities.
\item Encrypted transport for all inter-segment communication, with TLS inspection where organisational policy and legal requirements permit.
\end{itemize}

\subsection{Domain 3: Data Protection and Classification}

The Bill's customer notification requirements and supply chain provisions demand that organisations understand where regulated data resides, how it flows, and who can access it.

\textbf{Architectural requirements:}
\begin{itemize}
\item Data classification scheme mapped to CS\&R significance thresholds, identifying data whose compromise would constitute a qualifying incident.
\item Data flow mapping across organisational boundaries, including MSP-accessible data stores, supply chain data exchanges, and data centre tenant boundaries.
\item Data loss prevention (DLP) controls at segment boundaries, informed by classification and aligned to access policy.
\item Encryption at rest and in transit for all data classified as material to essential service delivery.
\end{itemize}

\subsection{Domain 4: Security Operations and Detection}

The 24/72-hour reporting regime is the Bill's most operationally demanding provision and requires a Security Operations Centre (SOC) architecture designed for speed, comprehensiveness, and regulatory integration.

\textbf{Architectural requirements:}
\begin{itemize}
\item Comprehensive telemetry collection from all components of the network and information systems relied upon for the essential service, including endpoint, network, identity, application, and cloud infrastructure layers.
\item Centralised Security Information and Event Management (SIEM) with retention periods aligned to regulatory and forensic requirements (minimum 12 months recommended).
\item Security Orchestration, Automation and Response (SOAR) with pre-built playbooks for CS\&R qualifying incident classification, notification workflow, and evidence packaging.
\item Automated significance classification: rules and models that evaluate detected events against the Bill's significance thresholds and escalate appropriately.
\item Integration with NCSC reporting interfaces and sector competent authority notification channels.
\end{itemize}

\subsection{Domain 5: Governance, Risk, and Compliance Architecture}

The Bill's continuous assurance model, cost recovery provisions, and CAF alignment requirements demand a GRC architecture that produces compliance evidence as a by-product of normal operations rather than through periodic manual assessment.

\textbf{Architectural requirements:}
\begin{itemize}
\item Continuous control monitoring: automated assessment of security control effectiveness, mapped to CAF objectives and CS\&R duties.
\item Risk register integration: dynamic risk scoring that incorporates real-time threat intelligence, vulnerability data, and supply chain risk assessments.
\item Regulatory reporting automation: structured data exports for competent authorities, supporting cost recovery transparency and compliance evidence.
\item Board reporting integration: automated dashboards providing directors with the oversight required under the Cyber Governance Code of Practice~\cite{ukgov2026cybergov} and the Board Toolkit~\cite{ncscboard2024}.
\end{itemize}

\section{Supply Chain Security Architecture}
\label{sec:supply}

The CS\&R Bill's supply chain provisions---the critical supplier designation power, MSP regulation, and supply chain risk management duties---demand an architectural approach to third-party security that goes well beyond contractual requirements and annual questionnaires.

\subsection{The Supply Chain Trust Boundary Problem}

In a perimeter-centric model, the trust boundary coincides with the network perimeter: everything inside is trusted, everything outside is untrusted. The reality of modern service delivery---particularly in sectors with extensive MSP dependencies---is that supply chain partners operate \emph{inside} the trust boundary by necessity. An MSP managing an organisation's cloud infrastructure has administrative access to the very systems that deliver the essential service. A pathology provider connected to NHS systems processes clinical data that is integral to healthcare delivery.

The Bill's provisions implicitly recognise this reality by regulating MSPs directly and empowering regulators to designate critical suppliers. But the architectural implication is that organisations must redesign their trust model to treat supply chain access as \emph{always untrusted until verified}---the core Zero Trust principle applied to inter-organisational relationships.

\subsection{Architectural Patterns for Supply Chain ZTA}

\textbf{Pattern 1: Brokered access with session-level authorisation.} All MSP and supplier access is mediated through a policy enforcement point (PEP) that evaluates identity, device posture, and contextual risk before granting time-bounded, least-privilege access to specific resources. No persistent network access is granted.

\textbf{Pattern 2: Mutual attestation.} Both the organisation and its supplier continuously attest to their security posture through automated mechanisms (e.g., device certificates, compliance signals, vulnerability scan results). Access policies dynamically adjust based on the supplier's attested posture.

\textbf{Pattern 3: Segregated operational environments.} MSP activities are confined to dedicated, monitored segments with full session recording and anomaly detection. Administrative actions in these segments are logged to an immutable store outside the MSP's control.

\textbf{Pattern 4: Supply chain risk scoring.} An automated system aggregates supplier security signals---attestation data, incident history, public vulnerability disclosures, regulatory status---into a dynamic risk score that informs both access policy and regulatory reporting.

These patterns are not mutually exclusive; a mature implementation will combine all four. The critical supplier designation power means that organisations must be architecturally prepared for the possibility that any supplier could be brought into scope at regulatory notice, requiring pre-existing integration points for assurance data collection and access governance.

\subsection{Alignment with Financial Services Frameworks}

Organisations in UK financial services face the additional challenge of aligning supply chain architecture with the FCA/PRA Critical Third Parties regime~\cite{fca2024ct3p} and the PRA's Supervisory Statement SS2/21 on outsourcing and third-party risk management~\cite{pra2024ss2_21}. The CS\&R Bill's critical supplier provisions are conceptually similar to the financial services critical third parties framework, and a unified architectural approach can satisfy both regimes simultaneously. The key is to design the supply chain governance architecture once, with mappings to multiple regulatory outputs.

\section{Incident Detection and Reporting Architecture}
\label{sec:incident}

\subsection{The 24/72-Hour Architecture}

The CS\&R Bill's incident reporting regime is the most operationally demanding provision from an architectural perspective. It requires two distinct capabilities: the ability to \emph{detect} qualifying incidents within a timeframe that permits 24-hour notification, and the ability to \emph{investigate and report} with sufficient detail within 72 hours.

\subsubsection{Detection Architecture}

Meeting the 24-hour notification requirement means that the combined time for detection, initial triage, and classification must be well under 24 hours---a practical target of 12 hours or less for mean-time-to-detect (MTTD) for qualifying incidents. This requires:

\textbf{Telemetry completeness.} Every component of the network and information systems relied upon for the essential service must generate security-relevant telemetry. Gaps in telemetry coverage are gaps in detection coverage, and therefore gaps in compliance. This is an architectural decision, not an operational one: telemetry must be designed into systems from deployment, not retrofitted.

\textbf{Detection engineering.} Detection logic must be calibrated to the Bill's significance thresholds. The shift from ``significant disruption'' to ``potential to cause significant impact'' means that detection rules must capture a broader category of events, including precursor activity such as unauthorised access attempts, lateral movement, and data staging. This requires a layered detection approach combining signature-based detection, behavioural analytics, and threat intelligence correlation.

\textbf{Automated classification.} A triage automation layer that applies the Bill's significance criteria to detected events and routes qualifying incidents directly into the notification workflow, reducing the human bottleneck in the classification process.

\subsubsection{Reporting Architecture}

The 72-hour full report demands forensic readiness: the ability to rapidly reconstruct the timeline, scope, and impact of an incident from preserved evidence.

\textbf{Immutable logging.} Security logs must be written to an immutable store (write-once, read-many) to ensure forensic integrity. This is particularly important where MSPs or suppliers have administrative access---logs must be outside their ability to modify or delete.

\textbf{Structured incident data model.} A standardised data model for incident records that maps directly to the reporting fields expected by competent authorities and the NCSC, enabling automated or semi-automated report generation.

\textbf{Evidence preservation.} Architectural capability to capture and preserve forensic evidence (memory dumps, disk images, network captures) without disrupting service delivery---a particular challenge in OT environments.

\subsection{Customer Notification Architecture}

The Bill requires MSPs, data centres, and digital service providers to notify affected customers of qualifying incidents. This creates an architectural requirement for:

\textbf{Impact blast radius mapping.} The ability to determine, within the notification timeframe, which customers are affected by a given incident. This requires maintained, accurate mappings between infrastructure components, services, and customer tenancies.

\textbf{Automated notification.} Pre-built notification workflows that can be triggered from the incident management system, with templates tailored to different customer categories and incident types.

\section{Cross-Regulatory Alignment}
\label{sec:crossreg}

\subsection{The Convergence Challenge}

UK organisations operating across regulated sectors face a growing web of overlapping cyber security obligations. The CS\&R Bill adds to, rather than replaces, existing requirements under sector-specific legislation such as the Telecommunications Security Act 2021~\cite{tsa2021} and the Financial Services and Markets Act 2023~\cite{fsma2023}. For UK financial services firms with EU operations, the simultaneous application of CS\&R, DORA~\cite{eu2022dora}, and NIS2~\cite{eu2022nis2} creates a complex compliance landscape.

\subsection{The Dual-Compliance Architecture}

The companion paper~\cite{shelby2025csr} establishes a ``highest common denominator'' approach to dual compliance: where CS\&R and DORA requirements overlap, adopt the higher standard as the baseline across the organisation. This architectural strategy avoids the cost and complexity of maintaining parallel compliance programmes.

From an architectural perspective, the key convergence points are:

\textbf{Incident reporting.} DORA requires a 4-hour initial notification for major ICT-related incidents, followed by intermediate reports within 72 hours and a final report within one month. The CS\&R Bill requires 24-hour initial notification and 72-hour full report. The architectural implication is clear: design to the 4-hour DORA standard and the CS\&R requirement is automatically satisfied. The detection architecture described in Section~\ref{sec:incident} should target a 4-hour MTTD where DORA applies.

\textbf{Third-party risk management.} DORA's ICT third-party risk management requirements are substantially more prescriptive than the CS\&R Bill's expected supply chain duties, covering contractual provisions, concentration risk, and the EU Critical ICT Third-Party Provider framework. The supply chain architecture described in Section~\ref{sec:supply}, designed to the DORA standard, will satisfy CS\&R requirements with minimal additional effort.

\textbf{Resilience testing.} DORA mandates threat-led penetration testing (TLPT) for significant financial entities, while the CS\&R Bill's testing expectations will be defined in secondary legislation but are expected to align with NCSC CAF Objective D (minimising the impact of incidents). Designing a unified resilience testing programme that satisfies both regimes is architecturally more efficient than maintaining separate testing schedules.

\textbf{Governance.} Both regimes emphasise board-level accountability for cyber risk. The architectural response is a unified governance dashboard that maps control status and risk posture to both CS\&R and DORA requirements, providing directors with a single view of compliance.

\subsection{NIS2 Alignment}

For UK organisations with EU operations subject to NIS2, the CS\&R Bill's provisions are substantially aligned with NIS2's requirements, reflecting their shared legislative heritage. The Bill's policy statement explicitly notes the intention to bring ``closer alignment with NIS2''~\cite{ukgov2025csrpolicy}. In architectural terms, an organisation designed to meet NIS2's requirements will find CS\&R compliance a relatively modest incremental effort. The principal areas of divergence---scope definitions, specific reporting timelines, and enforcement mechanisms---are regulatory rather than architectural in nature.

However, the alignment is not total, and CISOs should be aware of several practical divergences. NIS2's 24-hour ``early warning'' to the national CSIRT is followed by a 72-hour incident notification---broadly similar to the CS\&R regime---but NIS2 additionally requires a final report within one month containing a detailed description of the incident, root cause analysis, and cross-border impact assessment. The CS\&R Bill's reporting requirements in secondary legislation may or may not include a comparable final reporting obligation. Architecturally, the prudent approach is to design forensic investigation and root cause analysis capabilities that can support final reporting to both regimes.

NIS2 also introduces explicit requirements for supply chain security risk assessments at a national and Union level, and imposes management body accountability with personal liability provisions that go further than the current CS\&R Bill text. For UK organisations that must comply with both, the governance architecture should include management body training records, documented risk acceptance decisions, and personal accountability mapping that satisfies both regimes.

\subsection{Unified Compliance Architecture}

The convergence of CS\&R, DORA, and NIS2 creates a compelling case for a unified compliance architecture rather than parallel programmes. The architectural approach is to design a single security control framework that maps to all applicable regulatory requirements, with regulatory-specific reporting as an output layer rather than a design driver.

Practically, this means maintaining a single control catalogue (based on NCSC CAF v4.0 or ISO 27001~\cite{iso27001} controls, augmented for DORA-specific requirements), a single risk register with multi-regulatory impact assessment, a single incident management process with configurable notification workflows for different regulatory channels, and a single supply chain assurance framework with regulatory-specific reporting outputs. This unified approach reduces operational overhead, eliminates the risk of conflicting control implementations, and provides the CISO with a single view of the organisation's compliance posture across all applicable regimes.

\section{NCSC CAF v4.0 Alignment}
\label{sec:caf}

The NCSC Cyber Assessment Framework v4.0~\cite{ncsc2025caf4} is the government's preferred assurance framework for NIS-regulated organisations and is expected to be placed on a ``firmer footing'' under the CS\&R Bill's reforms~\cite{ukgov2026factsheets}. The reference architecture proposed in this paper maps to the CAF's four objectives as follows.

\textbf{Objective A: Managing security risk.} The GRC architecture (Domain 5) provides continuous risk assessment, board-level governance integration, and supply chain risk management. The identity governance architecture (Domain 1) supports the CAF's requirements for understanding assets, managing identities, and controlling access.

\textbf{Objective B: Protecting against cyber attack.} The network segmentation architecture (Domain 2) and data protection architecture (Domain 3) directly address Objective B's requirements for service protection policies, identity and access control, data security, and system security. Micro-segmentation and ZTNA implementation provide the architectural foundation for CAF indicators B2 (Identity and Access Control), B3 (Data Security), and B4 (System Security).

\textbf{Objective C: Detecting cyber security events.} The security operations architecture (Domain 4) is designed specifically to meet CAF Objective C's requirements for security monitoring and anomaly detection. The comprehensive telemetry, centralised SIEM, and automated classification capabilities align directly with CAF indicators C1 (Security Monitoring) and C2 (Proactive Security Event Discovery).

\textbf{Objective D: Minimising the impact of incidents.} The incident detection and reporting architecture (Section~\ref{sec:incident}), forensic readiness capabilities, and supply chain incident communication architecture map to CAF indicators D1 (Response and Recovery Planning) and D2 (Lessons Learned). The immutable logging and evidence preservation capabilities support effective post-incident analysis.

This mapping demonstrates that the proposed reference architecture provides a coherent technical foundation for CAF compliance, not as a separate compliance exercise but as an inherent property of the architecture itself.

\section{Maturity-Based Adoption Roadmap}
\label{sec:roadmap}

Recognising that organisations begin from different starting points, this section presents a three-phase adoption roadmap aligned to the CISA Zero Trust Maturity Model~\cite{cisa2023ztmm} levels and calibrated to the CS\&R Bill's phased implementation timeline.

\subsection{Phase 1: Foundation (0--12 Months Post-Royal Assent)}

This phase targets the ``Traditional'' to ``Initial'' maturity transition and focuses on establishing the minimum viable architecture for CS\&R compliance.

\textbf{Priority actions:}
\begin{itemize}
\item Conduct a comprehensive asset inventory of all network and information systems supporting essential service delivery.
\item Implement or enhance centralised log aggregation to achieve telemetry coverage across all critical systems.
\item Deploy or upgrade SIEM with detection rules calibrated to CS\&R qualifying incident thresholds.
\item Establish privileged access management (PAM) for all MSP and supplier administrative access, with session recording.
\item Develop incident response playbooks aligned to the 24/72-hour reporting regime, including templates and notification workflows.
\item Map existing security controls to NCSC CAF v4.0 objectives and identify gaps.
\item Conduct a supply chain dependency analysis to identify potential critical supplier designations.
\end{itemize}

\subsection{Phase 2: Segmentation and Integration (12--24 Months)}

This phase targets the ``Initial'' to ``Advanced'' maturity transition and focuses on implementing the core ZTA architectural patterns.

\textbf{Priority actions:}
\begin{itemize}
\item Implement micro-segmentation between critical system zones, MSP-accessible segments, and general corporate infrastructure.
\item Deploy ZTNA or SDP for all third-party and remote access, replacing VPN-based full-network access.
\item Implement continuous identity verification with device posture assessment and contextual risk scoring.
\item Deploy automated supply chain risk scoring and continuous third-party assurance mechanisms.
\item Integrate SOAR platform with detection, classification, and regulatory reporting workflows.
\item Implement data classification and DLP at segment boundaries for data material to essential services.
\item Establish continuous control monitoring with automated compliance evidence generation.
\end{itemize}

\subsection{Phase 3: Optimisation and Resilience (24--36 Months)}

This phase targets the ``Advanced'' to ``Optimal'' maturity transition and focuses on architectural optimisation, advanced capabilities, and demonstrated resilience.

\textbf{Priority actions:}
\begin{itemize}
\item Implement full ABAC-driven policy enforcement across all access decisions, including API-level controls.
\item Deploy advanced behavioural analytics and machine learning-based anomaly detection.
\item Establish mutual attestation with critical suppliers, with dynamic access policy adjustment based on supplier posture.
\item Conduct threat-led penetration testing (aligned to DORA TLPT where applicable) against the ZTA architecture.
\item Implement architectural agility capability: the ability to rapidly implement directed security changes in response to Secretary of State powers of direction.
\item Achieve full CAF v4.0 alignment with automated continuous assessment.
\item Develop board-level governance dashboards integrating CS\&R, DORA, and NIS2 compliance status.
\end{itemize}

\subsection{Resource and Investment Considerations}

The CS\&R Bill's impact assessment estimates total costs of less than \pounds150 million per year across all regulated entities~\cite{ukgov2026factsheets}. Individual organisations should anticipate costs varying significantly based on current maturity, the number of essential services in scope, and the complexity of supply chain dependencies. The phased approach proposed above allows organisations to prioritise investment against the highest-risk gaps identified in Phase 1 and spread capital expenditure over the implementation timeline.

CISOs should present the business case not as a compliance cost but as an investment in architectural capability that simultaneously satisfies CS\&R, supports DORA compliance for financial services, aligns with CAF v4.0, and---critically---reduces the operational risk of breach, which the Bill's incident reporting requirements make far more visible and consequential than under the current regime.

\subsection{The CISO's Strategic Role}

The CS\&R Bill fundamentally changes the CISO's position within the organisation. The enhanced enforcement regime (penalties up to \pounds17.5 million or 4\% of global turnover), the mandatory incident reporting obligations, and the Secretary of State's powers of direction create a regulatory environment in which cyber security failures have board-level consequences. This is reinforced by the Cyber Governance Code of Practice~\cite{ukgov2026cybergov}, which sets explicit expectations for board-level engagement with cyber risk.

For CISOs, this means that the architectural transformation described in this paper must be positioned as a strategic business programme, not a technology initiative. The key elements of this strategic positioning are:

\textbf{Board engagement.} CISOs must articulate the CS\&R Bill's architectural implications in business terms: the risk of non-compliance (financial penalties, regulatory intervention, reputational damage), the operational risk of incidents that the current architecture cannot detect within mandated timeframes, and the competitive advantage of demonstrable security maturity in supply chain relationships. The reference architecture's governance domain (Domain 5) provides the reporting framework for this engagement.

\textbf{Cross-functional ownership.} CS\&R compliance is not a security-only programme. The Bill's supply chain provisions require procurement and vendor management involvement. Incident reporting requires legal and communications readiness. The expanded scope definitions require business units to identify which network and information systems support essential service delivery. The CISO must orchestrate this cross-functional effort, using the architectural framework as the integrating structure.

\textbf{Regulatory relationship management.} The Bill's twelve competent authorities will have varying levels of maturity and expectations. CISOs should engage proactively with their sector regulator, using the NCSC CAF v4.0 mapping presented in Section~\ref{sec:caf} as a shared language for demonstrating architectural progress. Early engagement is particularly important for organisations that may be newly in scope (MSPs, data centres) and have no existing regulatory relationship.

\textbf{Budget and resourcing strategy.} The phased roadmap provides a basis for multi-year investment planning. Phase 1 priorities (telemetry, SIEM, PAM, incident playbooks) represent relatively bounded investments with clear compliance outcomes. Phase 2 investments (micro-segmentation, ZTNA, supply chain integration) are larger but can be justified against the dual CS\&R/DORA business case for financial services firms. Phase 3 investments (advanced analytics, mutual attestation, architectural agility) deliver the maturity that distinguishes genuine resilience from minimum compliance.

\section{Conclusion}
\label{sec:conclusion}

The UK Cyber Security and Resilience Bill is not merely a regulatory update; it is an architectural forcing function. Its provisions---expanded scope to MSPs, data centres, and critical suppliers; mandatory 24/72-hour incident reporting; supply chain security duties; Secretary of State powers of direction; and enhanced enforcement---collectively create requirements that perimeter-centric security architectures cannot satisfy.

This paper has demonstrated that Zero Trust Architecture provides the most coherent technical foundation for CS\&R compliance, not because ZTA is a universal solution to all security challenges, but because the Bill's specific requirements---continuous verification, granular access control, comprehensive monitoring, supply chain trust management, and architectural agility---map directly to ZTA's foundational principles.

The reference architecture and maturity-based roadmap presented here provide CISOs and security architects with an actionable framework for translating the Bill's regulatory requirements into architectural decisions. The key recommendations are:

\textbf{First}, treat the CS\&R Bill as an architectural transformation programme, not a compliance project. The Bill's provisions cannot be satisfied by adding controls to an existing perimeter-centric architecture; they require rethinking the trust model, segmentation strategy, detection architecture, and supply chain integration patterns.

\textbf{Second}, design for the highest common denominator across all applicable regulatory regimes. Organisations subject to CS\&R, DORA, and NIS2 should adopt the most demanding requirement in each domain as their architectural baseline. This approach is more expensive upfront but dramatically reduces the long-term cost and complexity of maintaining parallel compliance programmes.

\textbf{Third}, invest in detection and reporting architecture now. The 24-hour notification requirement is the Bill's most operationally demanding provision and requires the longest lead time to implement effectively. Organisations that wait for Royal Assent to begin building detection capabilities will find themselves non-compliant at enforcement.

\textbf{Fourth}, embed supply chain security into the architecture, not the contract. Contractual security requirements are necessary but insufficient; the Bill's supply chain provisions demand architectural controls that provide continuous visibility into third-party access, activity, and risk posture.

\textbf{Fifth}, use the NCSC CAF v4.0 as the assurance backbone. The CAF provides a structured, well-understood framework for demonstrating that the architecture meets the Bill's security and resilience requirements. Designing the architecture with CAF traceability from the outset simplifies the assurance process and supports regulatory engagement.

\textbf{Sixth}, anticipate secondary legislation. The Bill's primary legislation establishes the framework, but the operational detail---specific security requirements, incident reporting formats, critical supplier designation criteria, and cost recovery mechanisms---will be defined in secondary legislation following consultation expected in 2026. CISOs should engage with the consultation process and design their architecture with sufficient flexibility to accommodate the specific requirements that emerge. The reference architecture proposed in this paper is deliberately capability-based rather than control-prescriptive, providing this flexibility.

\subsection{Looking Ahead}

Several developments will shape the CS\&R compliance landscape over the coming 12--24 months and merit attention from security architects.

The \textbf{secondary legislation consultation}, expected in 2026, will define the specific security and resilience requirements that give operational effect to the Bill's framework. Security architects should monitor this process closely, as the technical detail will determine the precise calibration of detection thresholds, reporting formats, and supply chain assurance expectations.

The \textbf{NCSC CAF v4.0 adoption} across regulated sectors will establish the practical standard against which architectural compliance is assessed. Organisations that align their architecture to CAF v4.0 now will be ahead of the curve; those that wait for formal regulatory guidance risk a compressed implementation timeline.

The \textbf{critical supplier designation process} will reveal which suppliers regulators consider systemically important. Organisations should conduct their own dependency analysis now---the supply chain architecture patterns described in Section~\ref{sec:supply} should be in place before any designation occurs, not in response to it.

The \textbf{cross-border regulatory coordination} between the CS\&R Bill, DORA, and NIS2 regimes will evolve as all three mature. The security and defence partnership between the UK and EU, agreed in May 2025, includes provisions for cyber dialogue that may influence how the regimes converge or diverge in practice.

Finally, the \textbf{threat landscape} will continue to evolve. The Bill's powers of direction and future-proofing provisions are designed to enable rapid regulatory response to emerging threats. Organisations whose architecture is designed for agility---API-driven policy enforcement, modular security controls, automated response playbooks---will be better positioned to respond to directed requirements without destabilising their operational environment.

The organisations that begin architectural transformation now---in advance of Royal Assent and the secondary legislation that will define specific implementation requirements---will be best positioned not only for compliance but for genuine resilience against the escalating cyber threat to UK essential services.

\section*{Acknowledgements}

The author gratefully acknowledges the support and supervision of Professor Andrew Martin at the Department of Computer Science, University of Oxford. This work is informed by the author's experience in enterprise security architecture across UK financial services institutions and extends a companion practitioner guide to the CS\&R Bill~\cite{shelby2025csr}.

\begingroup
\raggedright
\sloppy
\bibliographystyle{ieeetr}
\bibliography{references}
\endgroup

\end{document}